\documentclass[]{spie}  %>>> use for US letter paper
\usepackage{amsmath,amsfonts,amssymb}
\usepackage{graphicx}
\usepackage[colorlinks=true, allcolors=blue]{hyperref}
\usepackage{subfig}

\title{First flight of the Gamma-Ray Imager/Polarimeter for Solar flares (GRIPS) instrument}

\author[a,b]{Nicole Duncan}
\author[b]{P. Saint-Hilaire}
\author[c]{A. Y. Shih}
\author[b]{G. J. Hurford}
\author[b]{H. M. Bain}
\author[d]{M. Amman}
\author[b]{B. A. Mochizuki}
\author[b]{J. Hoberman}
\author[b]{J. Olson}
\author[e,b]{B. A. Maruca}
\author[b]{N. M. Godbole}
\author[f]{D. M. Smith}
\author[g]{J. Sample}
\author[b]{N. A. Kelley}
\author[b]{A. Zoglauer}
\author[h]{A. Caspi}
\author[i]{P. Kaufmann}
\author[b]{S. Boggs}
\author[b]{R. P. Lin}

\affil[a]{UC Berkeley, 366 Leconte Hall, Berkeley, CA, 94720, USA}
\affil[b]{Space Sciences Laboratory, UC Berkeley, 7 Gauss Way, Berkeley, CA, 94720, USA}
\affil[c]{Heliophysics Science Division, NASA Goddard SFC, Greenbelt, MD, 20771, USA}
\affil[d]{Lawrence Berkeley National Laboratory, 1 Cyclotron Road, Berkeley, CA, 94720, USA}
\affil[e]{Dept of Physics and Astronomy, Univ. of Delaware, 4 Kent Way, Newark, DE, 19716, USA}
\affil[g]{Dept of Physics, UC Santa Cruz, 1156 High Street, Santa Cruz, CA, 95064, USA}
\affil[g]{Dept of Physics, EPS building, Montana State University, Bozeman, MT, 59717, USA}
\affil[h]{Southwest Research Institute, 1050 Walnut Street, Suite 300,
Boulder, CO, 80302, USA}
\affil[i]{Universidade Presbiteriana Mackenzie, Sao Paulo, SP, Brazil}

\authorinfo{Further author information: (Send correspondence to Nicole Duncan)\\Nicole Duncan: E-mail: nicoleduncan@berkeley.edu, Telephone: 1 510 642 1707\\  Pascal Saint-Hilaire: E-mail: pascal@ssl.berkeley.edu}

\begin{document} 
\maketitle

%%%%%% -----------------------------------------------------------------------------------

\begin{abstract}

The Gamma-Ray Imager/Polarimeter for Solar flares (GRIPS) instrument is a balloon-borne telescope designed to study solar-flare particle acceleration and transport. We describe GRIPS's first Antarctic long-duration flight in January 2016 and report preliminary calibration and science results.

Electron and ion dynamics, particle abundances and the ambient plasma conditions in solar flares can be understood by examining hard X-ray (HXR) and gamma-ray emission (20 keV to 10 MeV). Enhanced imaging, spectroscopy and polarimetry of flare emissions in this energy range are needed to study particle acceleration and transport questions. The GRIPS instrument is specifically designed to answer questions including: What causes the spatial separation between energetic electrons producing hard X-rays and energetic ions producing gamma-ray lines? How anisotropic are the relativistic electrons, and why can they dominate in the corona? How do the compositions of accelerated and ambient material vary with space and time, and why?

GRIPS's key technological improvements over the current solar state of the art at HXR/gamma-ray energies, the Reuven Ramaty High Energy Solar Spectroscopic Imager (RHESSI), include 3D position-sensitive germanium detectors (3D-GeDs) and a single-grid modulation collimator, the multi-pitch rotating modulator (MPRM). The 3D-GeDs have spectral FWHM resolution of a few hundred keV and spatial resolution $<$1 mm$^3$.  For photons that Compton scatter, usually $\gtrsim$150 keV, the energy deposition sites can be tracked, providing polarization measurements as well as enhanced background reduction through Compton imaging. Each of GRIPS's detectors has 298 electrode strips read out with ASIC/FPGA electronics. In GRIPS's energy range, indirect imaging methods provide higher resolution than focusing optics or Compton imaging techniques. The MPRM grid-imaging system has a single-grid design which provides twice the throughput of a bi-grid imaging system like RHESSI. The grid is composed of 2.5 cm deep tungsten-copper slats, and quasi-continuous FWHM angular coverage from 12.5-–162 arcsecs are achieved by varying the slit pitch between 1–-13 mm. This angular resolution is capable of imaging the separate magnetic loop footpoint emissions in a variety of flare sizes. In comparison, RHESSI's 35-arcsec resolution at similar energies makes the footpoints resolvable in only the largest flares.

\end{abstract}

% Include a list of keywords after the abstract 

%%%%%% -----------------------------------------------------------------------------------
\section{INTRODUCTION}
\label{sec:intro}  

  %science overview / goals (0.5 pages)
  %Figure: usual Oct 28 image with the RHESSI/GRIPS beam-width comparison
  
The Gamma-Ray Imager/Polarimeter for Solar Flares (GRIPS) instrument is a balloon-borne solar observatory which flew on its first mission during the 2015/16 Antarctic summer season. GRIPS provides imaging (12.5--162 arcseconds), spectroscopy ($\sim$30kev to $\gtrsim$10MeV) and polarimetry of high energy emission during solar flares. Observations from the GRIPS instrument address questions concerning the acceleration and transport of particles during solar flares. GRIPS is a NASA Heliophysics Low Cost Access to Space mission led by Pascal Saint-Hilaire at UC Berkeley's Space Science Laboratory (SSL). Most of the telescope and its support systems were designed, built and tested at SSL with contributions from Lawrence Berkeley National Laboratory (LBNL), NASA Goddard Space Flight Center, and UC Santa Cruz. 

\begin{figure} [!h]
\centering
\includegraphics[scale=0.04]{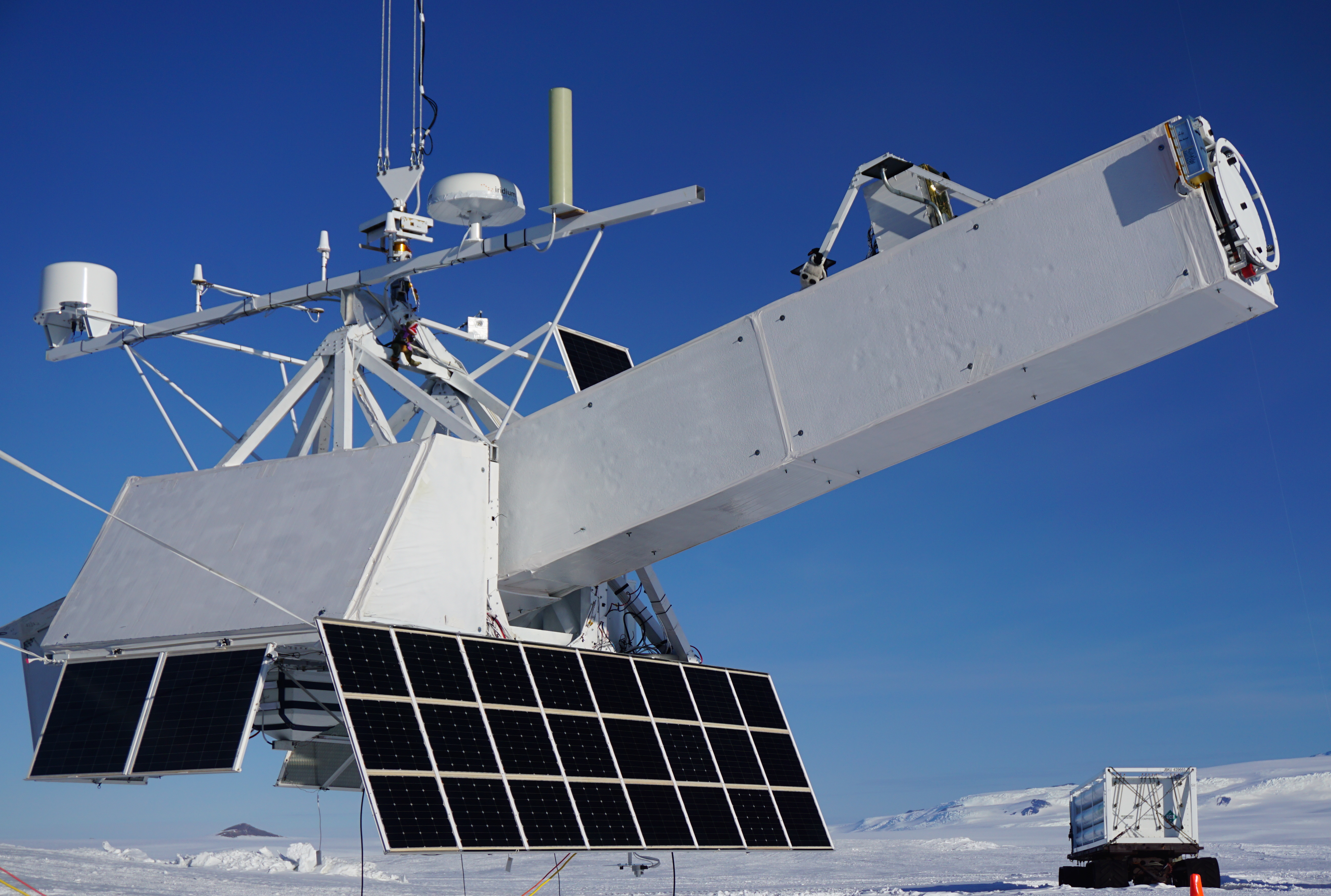} 
\caption{\label{fig:payload}The fully integrated GRIPS instrument payload. The gondola is suspended from "The Boss" launch vehicle while the GRIPS and Columbia Scientific Ballooning Facility (CSBF) teams complete pre-flight checklists during a launch attempt.}
\end{figure}

The GRIPS payload (Figure~\ref{fig:payload}) flew on a long-duration balloon over Antarctica on January 19--30, 2016. During GRIPS' $\sim$12-day flight, 21 C-class flares occurred. Because of where the payload landed and the lateness of the season, recovery resources were limited. All flight data for the primary instrument and its three piggyback instruments were recovered, but the instrument itself will remain in the Antarctic deep field during the 2016 winter and will be recovered in the following (2016/17) summer season. All systems performed in flight as designed. This paper reports on the details of GRIPS' first flight and provides updates on critical system changes and milestones since the previous GRIPS papers. \cite{Shih2012, Duncan2013}

\subsection{Science Goals}

Solar flares can release $10^{32}$--$10^{33}$ ergs within 10s--1000s of seconds, imparting up to tens of percent of the energy into particles\cite{Emslie2012}. During these events, electrons can be accelerated to 100s of MeV and ions to 10s of GeV. Large-scale field reconfigurations resulting from magnetic reconnection in the corona are thought to power flares, but the precise mechanisms that convert the stored magnetic energy into particle kinetic energy are poorly understood. 

Accelerated particles either escape into interplanetary space or rain down into the solar atmosphere where they lose energy as they interact with the increasing particle densities. The particles emit photons in a variety of processes\cite{Murphy2005, Murphy2007} which give clues to the underlying acceleration and transport mechanisms. High resolution imaging and spectroscopy of the hard X-ray/gamma-ray energy range during flares is required to build a complete picture of these processes. The Reuven Ramaty High Energy Solar Spectroscopic Imager (RHESSI)\cite{Lin2002}, launched in 2002, first opened the door to imaging and detailed spectroscopy in this energy range. 

\begin{figure} [!h]
\centering
\includegraphics[scale=0.6]{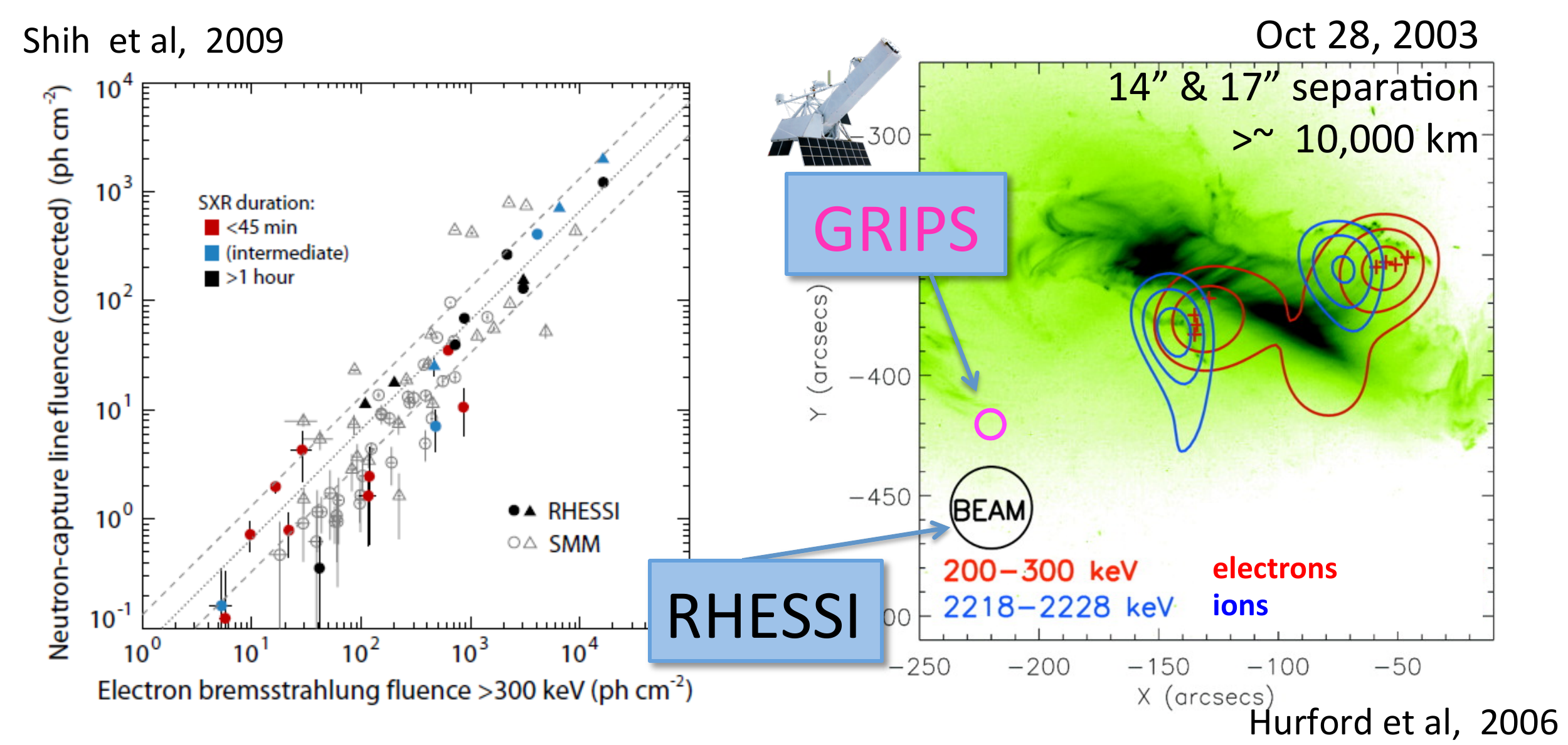} 
\caption{\label{fig:discrepancy}\emph{Left:} The correlation between the relativistic electron bremsstrahlung and the 2.2~MeV line fluence indicates that the same process may be accelerating protons and electrons. The correlation extends over 3 orders of magnitude and two different missions.\\
\emph{Right:} RHESSI gamma-ray image in two different energy bands of the GOES class X17 flare on October 28, 2003, overlaid on a TRACE 195 {\AA} image. The electron-associated bremsstrahlung emission is shown in red contours, and the imaged 2.2 MeV line is shown in blue contours. The electron- and ion-associated centroids are separated from one another by $\sim$17" in each footpoint. Inset is a comparison of the GRIPS and RHESSI minimum beam widths. At gamma-ray energies RHESSI can only separate footpoints in the largest flares. The GRIPS resolution is fine enough to separate footpoints in a variety of flare sizes.}
\end{figure}

RHESSI observations linked the high-energy photon emission to observed spatial structures like magnetic fields, verifying the overall geometry of the standard flare model. In the RHESSI and SMM datasets ion and relativistic electron fluences were seen to be correlated, showing that flares accelerate ions and relativistic electrons proportionally\cite{Shih2009} and suggesting that these populations are accelerated together, and possibly by similar mechanisms. However, imaging of two of the largest flares of the last solar cycle show that the centroids of ion and relativistic electron emission were significantly displaced from one another\cite{Hurford2003, Hurford2006}. This result is surprising; ions and electrons that are accelerated in the same region are thought to be transported out of the acceleration area along the same field lines, implying that they would enter the chromosphere together and have similar emission source locations. Figure~\ref{fig:discrepancy} shows the correlated emission as well as the observed spatial displacement. Unfortunately RHESSI's reduced imaging capabilities at gamma-ray energies limits these observations to the few largest and best observed flares of the past solar cycle. 

RHESSI and other solar observatories have revealed a new set of flare particle questions, which GRIPS was designed to address:
\begin{itemize}
\item What caused the observed spatial separation between the sites of relativistic electron and ion energy deposition in flare footpoints? Does this separation occur in all flares?
\item Do all flares accelerate ions?
\item Are relativistic electrons isotropic or beamed in the corona? 
\item What is the composition/variance of accelerated material in space and time?
\end{itemize}

%%%%%% -----------------------------------------------------------------------------------
\section{Instrument outline and overview}
\label{sec:inst} 

\begin{figure} [!h]
\centering
\includegraphics[scale=0.6]{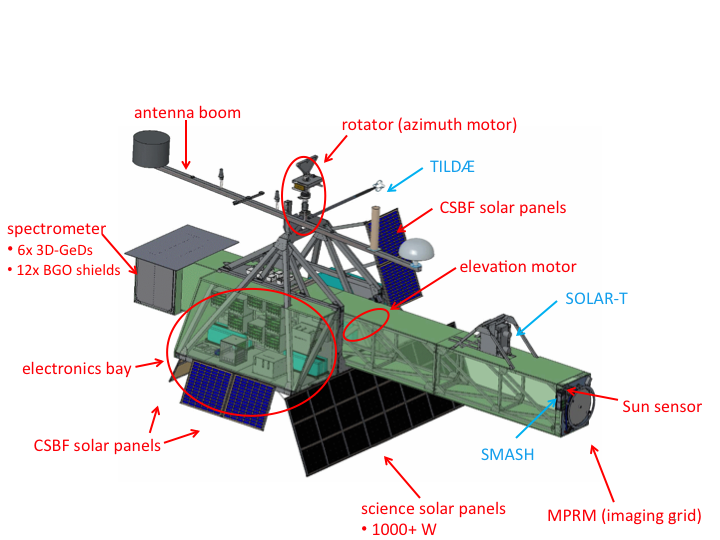} 
\caption{\label{fig:schematic}Schematic of the fully integrated GRIPS payload. The GRIPS and CSBF components are shown in red and the three piggyback instruments (see Section \ref{sec:piggyback}) are shown in blue. The flight computer, aspect computer, power distribution unit and the majority of the signal processing electronics are housed in the electronics bay. The electronics bay and telescope boom are both covered in opaque foam, but illustrated here as transparent to highlight interior components. Antennas were provided by CSBF.}
\end{figure}

The GRIPS instrument consists of a germanium spectrometer/polarimeter and a single-grid imager, separated by an eight-meter boom. GRIPS's key technological advances over the current HXR/gamma-ray solar state of the art (RHESSI) are 3D position-sensitive germanium detectors (3D-GeDs), and a coded-aperture-style imaging system, the multi-pitch rotating modulator (MPRM).  A detailed description of these components, the GRIPS science instrument overall and its support systems can be found in earlier manuscripts\cite{Shih2012,Duncan2013}. Figure \ref{fig:schematic} shows a schematic of the complete payload and its systems. In this section we give a general review of key systems, highlight instrumental milestones and discuss changes since we last reported on GRIPS\cite{Duncan2013}. 

The GRIPS spectrometer is able to resolve the site of individual photon interactions within a crystal in three dimensions. When photons Compton scatter in the crystal, which is the dominant process above $\sim$150 keV, GRIPS can resolve each energy deposition site and reconstruct the complete scatter path. Determining the scattering track enables polarization measurements and coarse Compton imaging ($\sim$1$^{\circ}$ resolution) that can be used to image extended sources or eliminate background.

\begin{figure} [!h]
\centering
\includegraphics[scale=.7]{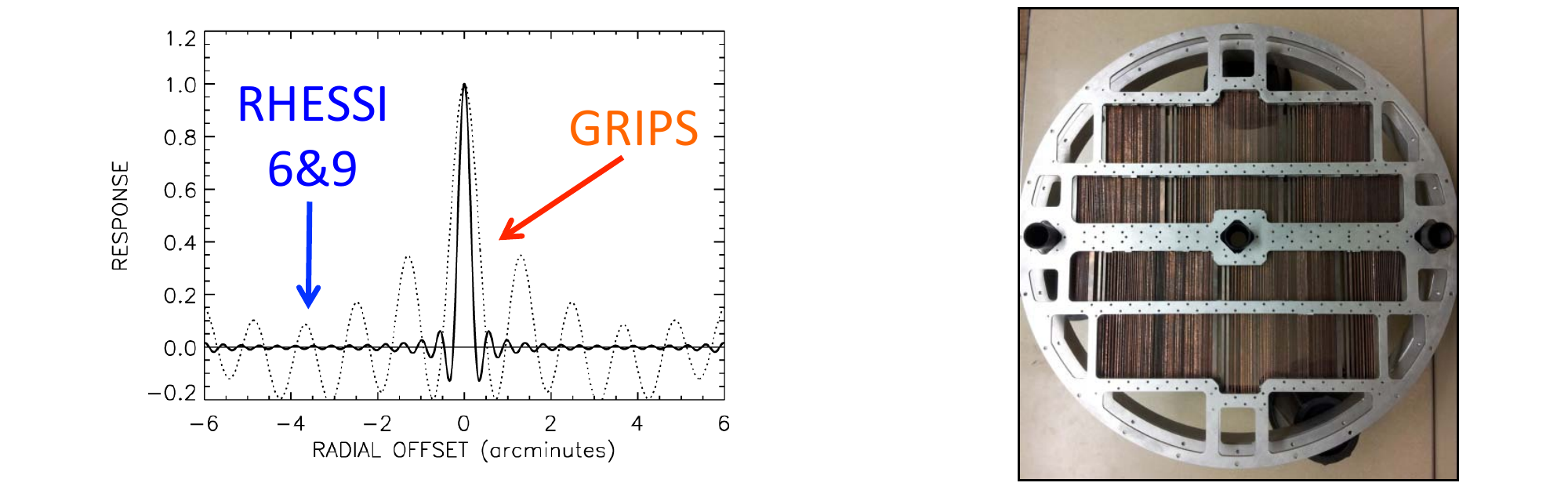} 
\caption{\label{fig:mask}\emph{Left:} The point response function of GRIPS's 13 spatial frequencies in comparison to RHESSI's two frequencies at gamma-ray energies. Note that GRIPS's response is virtually free of sidelobes.\\
\emph{Right:} The MPRM imaging grid contains a large number of parallel slats, each of which is made of a 70/30 tungsten/copper alloy that is 2.54 cm thick along the line-of-sight to the Sun. The slit pitch varies between 1--13 mm, quasi-continuously sampling between 12.5--162 arcsec resolution. Mounted along the grid diameter are three optical lenses that project images of the Sun as part of the aspect system (see Section \ref{sec:aspect}).}
\end{figure}

Detector position sensitivity is the enabling technology for GRIPS's single-grid imaging system.  The site of the first energy deposition for a given photon can be backprojected through the single grid to determine the possible origination regions on the Sun. This arrangement yields $\sim$2x greater photon throughput than related bi-grid designs, such as RHESSI, which had detectors with no position sensitivity. GRIPS's imaging grid has 13 slit pitches (see Figure \ref{fig:mask}), which rotate through 360$^{\circ}$ orientation at $\sim$10 rpm, providing quasi-continuous imaging resolution between 12.5--162 arcseconds. At gamma-ray energies, RHESSI's grids 6 \& 9 provide only 2 slit pitches, with a minimum angular resolution of $\sim$35 arcseconds. Quasi-continuous coverage in the UV plane imparts GRIPS with a nearly ideal point response function that is virtually free of side-lobes (Figure \ref{fig:mask}).

A single grid eases the tight engineering and manufacturing constraints of dual-grid designs, such as alignment and precision tolerances. This allows the 8-m boom to be free to twist and flex during flight. Though the grid position is not controlled, its location relative to the detectors is closely monitored by the GRIPS aspect system (see Section \ref{sec:aspect}). The aspect system provides high-precision monitoring of telescope pitch, yaw and roll. This enables arcsecond-precision aspect knowledge with a boom pointed to only 0.5$^{\circ}$ rms by the pointing control system (see Section \ref{sec:pointing}).

%\subsection{Timeline}
  %I want to take this out, unless someone feels strongly about keeping it

\subsection{Flight Cryostat}
\label{sec:cryostat}

\begin{figure} [!h]
\centering
\includegraphics[scale=0.6]{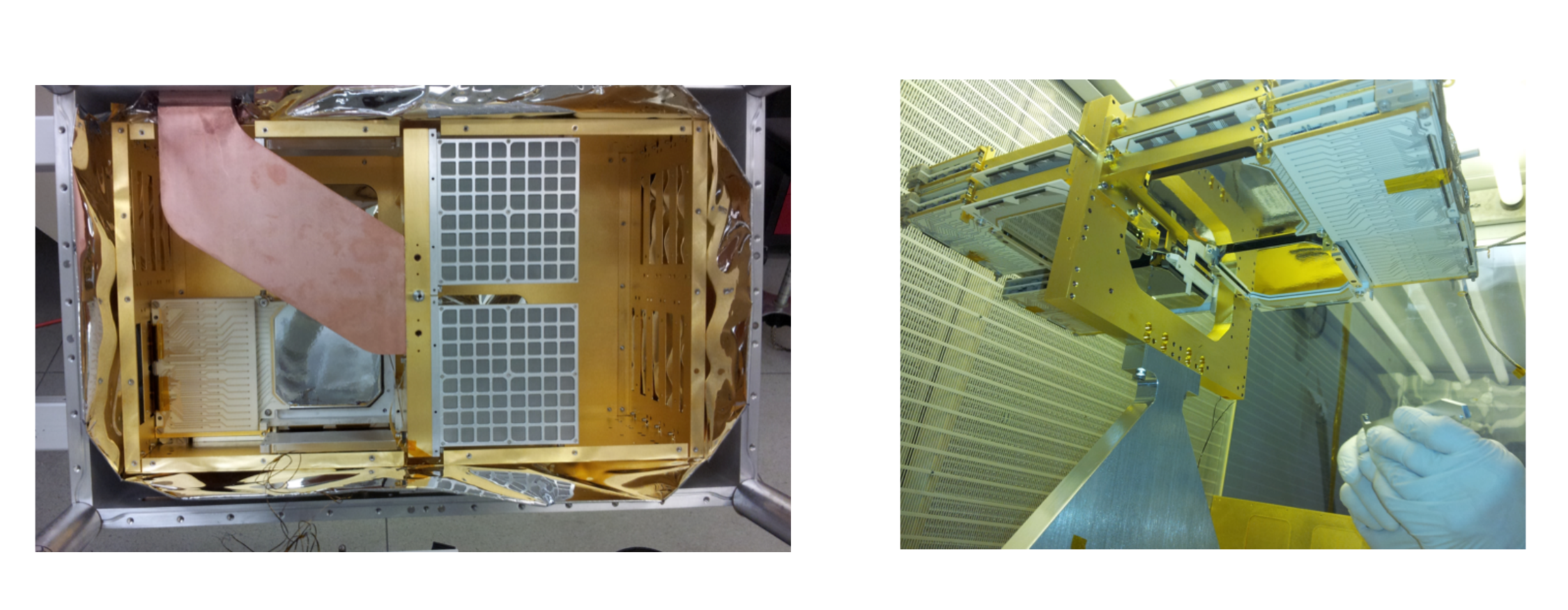} 
\caption{\label{fig:inside}\emph{Left:} The interior of the cryostat, integrated with two test detectors. The detector stacks are housed inside an infrared shield (gold) then wrapped in thermal blankets (silver) before being inserted into the aluminum cryostat shell. The detectors are cooled by a copper cold finger shown in the foreground.\\
\emph{Right:} The full flight detector stack.}
\end{figure}

The six 3D-GeDs are arranged in two planes, a 2$\times$2 plane in front of a 2$\times$1 plane, all housed within a single cryostat (Figure~\ref{fig:inside}). The detector stacks are attached to a copper coldfinger within the aluminum cryostat. The coldfinger is mechanically cooled by a Sunpower CryotelCT cryocooler which operated at $\sim$85 W to maintain an $\sim$80 K detector temperature during flight. Heat is passively rejected from the cryocooler through conduction to three radiator plates. To accommodate the wide range of ground operating conditions, including integration and final hang test in Texas where exterior temperatures can exceed 35$^{\circ}$C, an array of four thermoelectric coolers pumped heat from the cryocooler onto a finned heatsink with forced airflow.

The analog signals from GRIPS’s 1800 electrode strips pass through the cryostat wall via 48 polyamide flexible film circuits (.012 inch thickness) to accommodate the required signal density, with custom designed vacuum feedthroughs using Viton seals. Immediately outside the cryostat are 48 ASICs, 8 ASICs per detector, which handle triggering, shaping and A/D conversion via a Wilkinson ramp comparator. The GRIPS ASIC was designed by Gamma Medica IDEAS as an extension of the VATA450 ASIC family used on the ASTRO-H spacecraft \cite{Takahashi2012} and the FOXSI sounding rockets \cite{Krucker2013}. In Sections \ref{sec:detectors} and \ref{sec:spectra} we discuss the detectors and their performance.

%For two flex circuits, necessary modifications to the strain relief mechanism increased the circuit path length such that the end protruding from the cryostat was too short to mate with existing readout hardware. The signal traces on these two flex circuits were connected to ground. During the final stages of detector integration three flex circuits were found to have delaminated, causing vacuum leaks. For all 3 flex circuits, this required pushing the circuit back through the cryostat wall and disabling the channels by connecting them to ground.

The cryostat is surrounded by a 5-cm thick bismuth germanate (BGO) scintillator active anti-coincidence shield consisting of a hexagonal well surrounding the sides and bottom of the cryostat and read out by 36 individual PMTs. This shield was originally developed for the Max ‘91 HIgh REsolution Gamma-ray/hard X-ray Spectrometer (HIREGS), a balloon-borne predecessor to the RHESSI spacecraft, and has been successfully re-flown for NCT, and now GRIPS. With a threshold of $\sim$100 keV, the BGO shields significantly reduce the instrumental background even before the Compton imaging and the spectral signature techniques are applied.

%%%%%% -----------------------------------------------------------------------------------
\section{Plum Brook Thermal Vacuum Test}
\label{sec:PB} 

\begin{figure} [!h]
\centering
\includegraphics[scale=0.45]{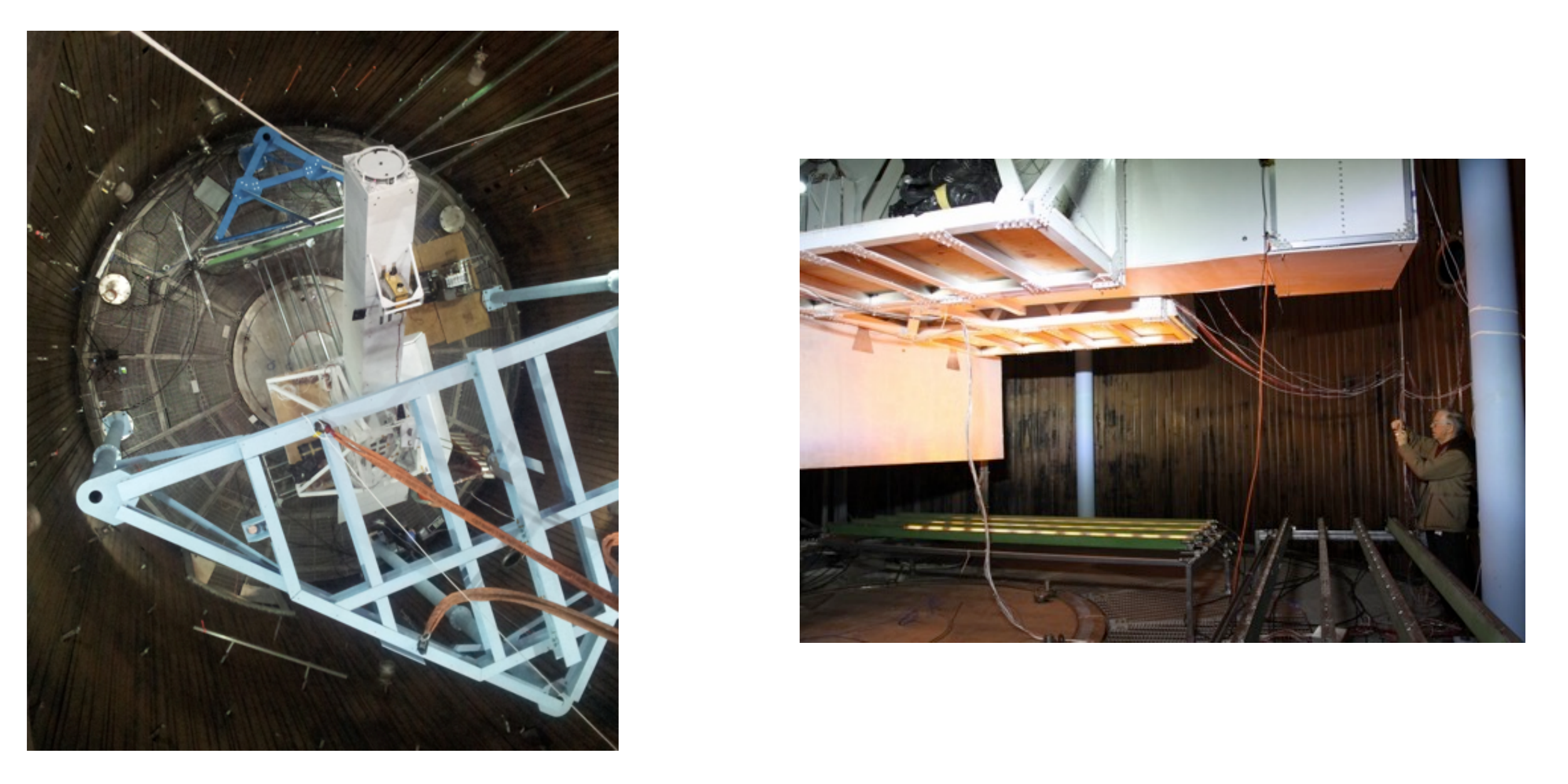} 
\caption{\label{fig:PB}Images of the GRIPS payload in the Plum Brook B-2 chamber. The blue support truss suspended the payload within the chamber and allowed for $\pm$5$^{\circ}$ pointing maneuvers. Calibrated IR lamps were used to simulate direct solar illumination and Earth's albedo.}
\end{figure}

It is critical to test subsystems to ensure that all mechanical and electrical systems survive the near-vacuum ($\sim$3 mbar) environment at stratospheric altitudes. In March 2015, the GRIPS gondola underwent a full instrument thermal vacuum test at NASA Glenn's Plum Brook Station to simulate the thermal environment at altitude. The test was carried out in the Spacecraft Propulsion Research Facility (B-2), which consists of a 33'$\times$120' chamber, and lasted 30 hours at $\sim$2 torr and 77K. The gondola was suspended from a specially built truss which allowed for $\pm$5$^{\circ}$ pointing maneuvers during the test (Figure \ref{fig:PB}).

The B-2 chamber's cold wall and lid were cooled with liquid nitrogen to simulate a space-like radiation environment. It is impossible to fully recreate flight-like conditions, i.e., a very intense point-source striking the front areas of the payload, and a large albedo source underneath. Instead, we used banks of calibrated IR lamps to simulate flight-like radiation on key areas of the payload. We simulated direct solar illumination with lamps placed (1) in front of the electronics bay and (2) in front of the boom. The Earth albedo lamps were placed (3) under the electronics bay and (4) under the spectrometer. Heaters were applied to some elements that were impractical to illuminate with IR lamps.

The GRIPS thermal model predictions reasonably matched the readings from the $\sim$80 temperature sensors used during the test. A critical parameter in making accurate predictions was properly accounting for the radiation environment of the B2 chamber's interior, namely that it cannot be modeled as a perfect black body due to degraded paint on the walls and the grated metal floor. Component temperatures in flight were in reasonable agreement with the overall GRIPS thermal model.

%%%%%% -----------------------------------------------------------------------------------
\section{Flight Campaign}
\label{sec:flight} 

%\begin{figure}[!h]
%\centering
%\includegraphics[scale=1]{gondola_for_antennas.jpg} 
%\caption{\label{fig:boss}The GRIPS gondola suspended from "The Boss" launch vehicle while on the flight line prior to launch.}
%\end{figure}

Beginning on October 31, 2015, the GRIPS gondola underwent a period of integration, pointing control and aspect system testing, and instrument calibration at the NASA Columbia Scientific Balloon Facility's (CSBF) Long Duration Balloon (LDB) camp on the Ross Ice Shelf near McMurdo Station, Antarctica. While most of the instrument was transported by ship to Antarctica, the more delicate GRIPS cryostat was flown by the United States Air National Guard (ANG) to McMurdo, stopping at several locations enroute. Due to logistical complications with requiring access to power to keep the cryostat cold, the cryostat was instead transported warm, at the expense of extending the period of integration in Antarctica by $\sim$2 weeks to cool the cryostat back down to $\sim$80 K.

After a pre-flight compatibility test, GRIPS was declared flight ready on the December 26, 2015. Over the following weeks, surface weather conditions needed for launch were scarce; in particular, even on clear days, strong surface winds exceeded the required 5 knot upper threshold for launch. During this time the payload had seven aborted launch attempts (``roll-outs'') before successfully launching on the January 19, 2016, at 01:41 UTC. Each roll-out consisted of a gondola pickup by CSBF's launch vehicle ``The Boss'', followed by a full communications and instrument checkout.% while suspended in a configuration shown in Figure \ref{fig:boss}
During several roll-outs, the payload was subjected to colder temperatures than it would encounter during flight, including freezing fog and light snow. 

\begin{figure} [!h]
\centering
\includegraphics[scale=.55]{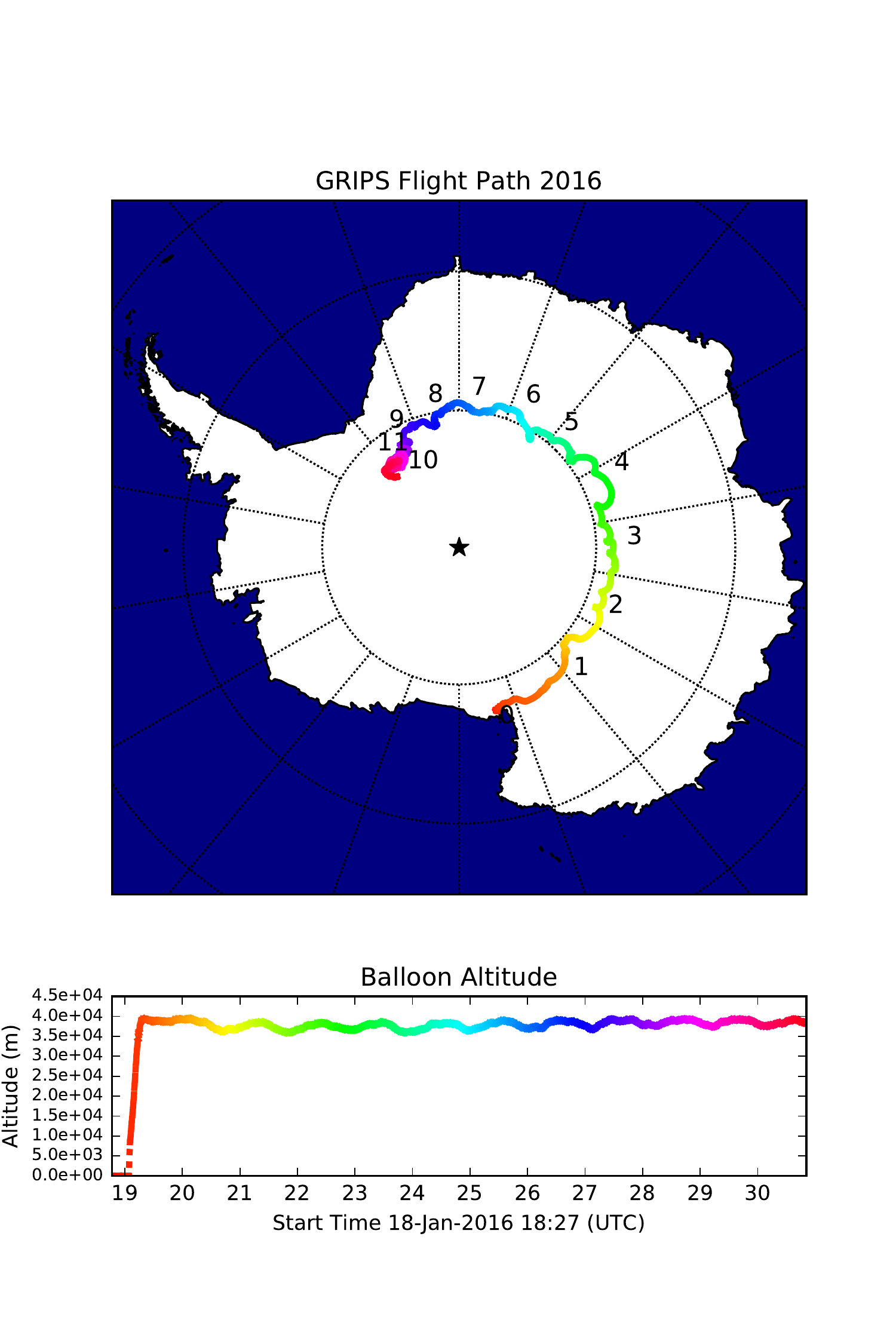} 
\caption{\label{fig:flightpath}\emph{Top:} GRIPS flight path with time represented by the color gradient progressing from red, at launch to blue on landing. Numerical annotation indicates the number of days at float.\\
\emph{Bottom:} Balloon altitude from launch with the color progression indicating time at float.}
\end{figure}

After $\sim$5 hours of ascent, the balloon reached a float altitude of $\sim$39,000 m (bottom of Figure \ref{fig:flightpath}). During the flight, the balloon altitude varied diurnally, reaching a maximum altitude of 39,459m at 03:06 UT on January 20 and a minimum altitude of 35,873 m at 20:08 UT on January 21. Figure \ref{fig:flightpath} (top) shows the balloon flight path around the Antarctic continent, with days at float annotated and represented in the color gradient (red at launch, blue/purple on landing). 

\begin{figure} [!h]
\centering
\includegraphics[scale=.7]{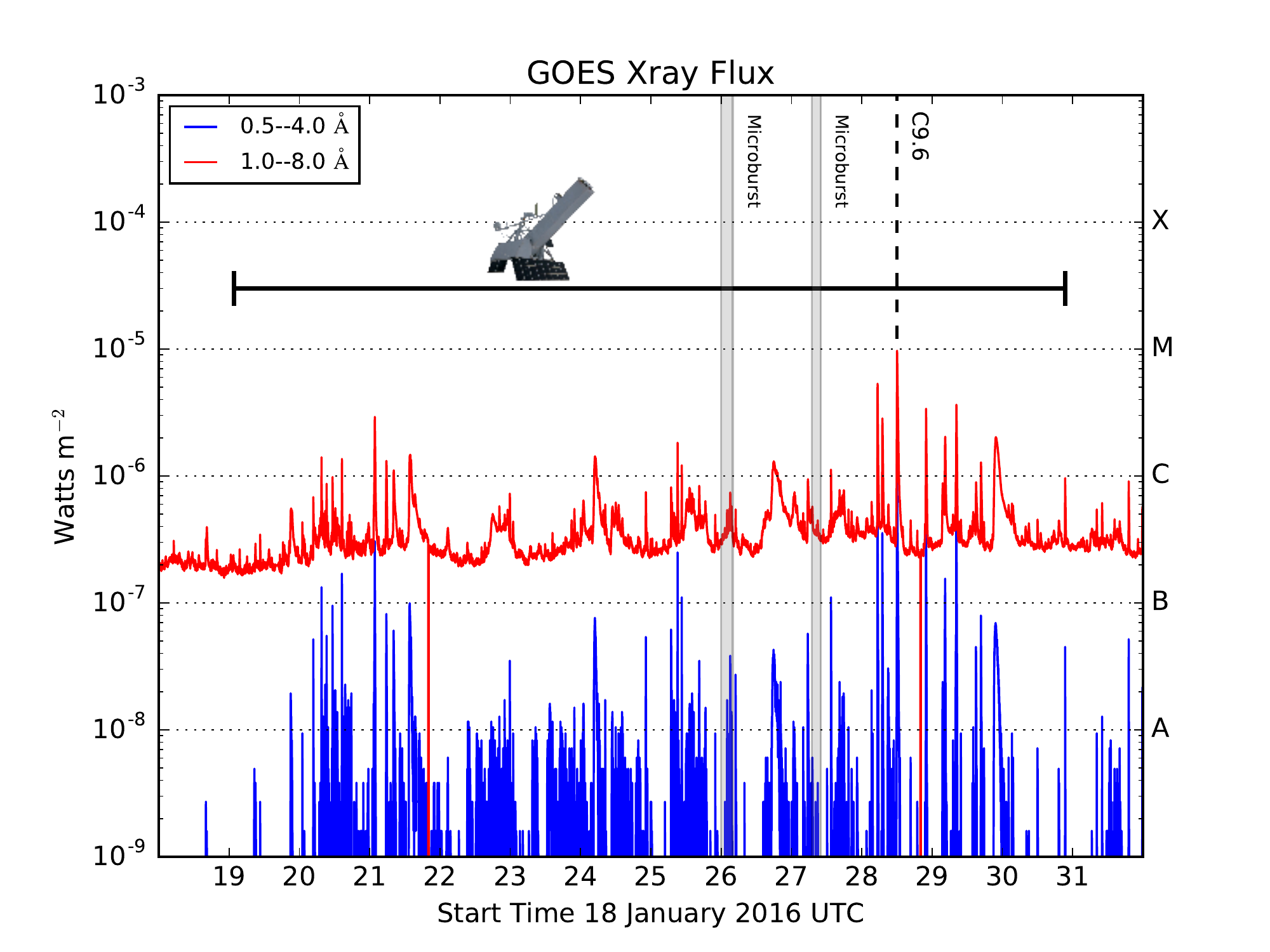} 
\caption{\label{fig:goes}GOES X-ray lightcurve for the duration of the GRIPS flight (marked by the horizontal black line). The vertical dashed line indicates the largest solar flare, GOES class C9.6, and the shaded areas indicate the periods of extended microburst activity.}
\end{figure}

Over the flight, solar activity was relatively low, producing 21 GOES C class flares, the largest of which was a C9.6 flare on January 28, 2016, at $\sim$12:00 UT. Figure \ref{fig:goes} shows the GOES X-ray lightcurves at 0.5--4.0 \AA{}(blue) and 1.0--8.0 \AA{}(red) for this time period. The corresponding GOES flare classifications are noted on the right-hand axis.  GRIPS also observed two extended periods of electrons precipitating from the radiation belts in events known as microbursts.

After nearly 12 days at float, preparations were made for landing. Detector data acquisition ended at 19:51 UT on January 30, with flight termination occurring shortly after, at 20:51 UT. The balloon landed at 21:30 UT in the Queen Maud Land region of the Antarctic continent, at $83^{\circ}6.18$ S, $40^{\circ}54.08$ W. Due to landing occurring late in the 2015/2016 Antarctic summer season, only the data vaults were recovered. The remainder of the instrument will be recovered in the following austral 2016/2017 summer.

%%%%%% -----------------------------------------------------------------------------------
\section{Detectors}
\label{sec:detectors}

%GRIPS uses high purity monolithic germanium cross-strip detectors to detect photon energy deposition in three dimensions. Above 150 keV photons are more likely to Compton scatter than have single site energy depositions. With GRIPS' 3D position resolution, the Compton scatter track can be reconstructed, yielding a measure of incident photon polarization and providing  coarse Compton-style imaging ($\sim$1$^{\circ}$ resolution) that can be used for background elimination and imaging extended sources. 

GRIPS's 3D-GeDs are high-purity germanium cross-strip detectors that can be used to locate and measure each energy deposition in three dimensions.  These detectors extend the virtually pixelated detector concept that was successfully demonstrated on the NCT missions in 2005\cite{Coburn2005} and 2009\cite{Bandstra2011} and on the COSI mission in 2016. The GRIPS detectors are of the same size (7.5~cm$\times$7.5~cm$\times$1.5~cm), but have a 4$\times$ finer (0.5~mm) electrode pitch. As with the earlier generations, these detectors were developed and fabricated at Lawrence Berkeley National Laboratory.

\begin{figure} [!h]
\centering
\includegraphics[scale=0.65]{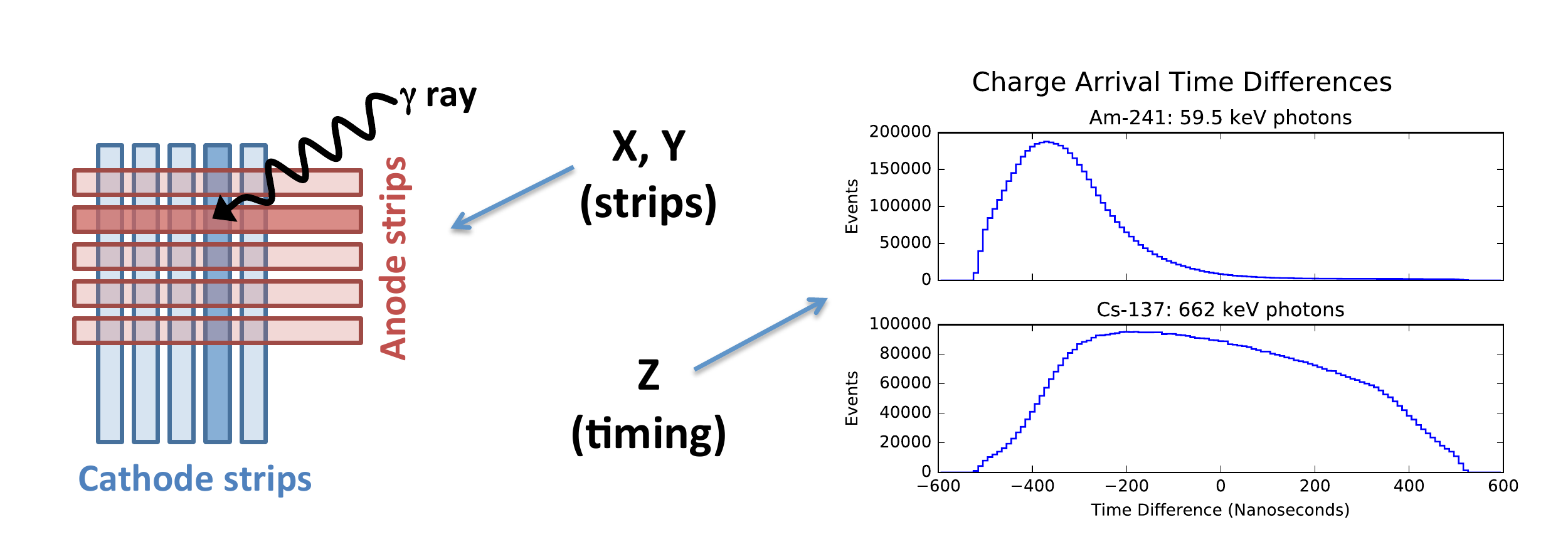} 
\caption{\label{fig:XYZ} \emph{Left:} A cartoon depicting how a 2D location (X, Y) is generated from coincident triggers on the anode and cathode strips.\\
\emph{Right:} Plots showing the time differences between charge collection on the anode and cathode for two different calibration sources, giving a measure of depth (Z). In this plot the anode trigger time is subtracted from the cathode trigger time. Energy deposition events that occur closer to the cathode (anode) are represented by negative (positive) values. For the Am-241 source, 59.5 keV  photons deposit close to the illuminated surface.  For the Cs-137 source, the 662 keV and continuum photons penetrate more uniformly through the crystal.}
\end{figure}

\begin{figure} [!h]
\centering
\includegraphics[scale=0.75]{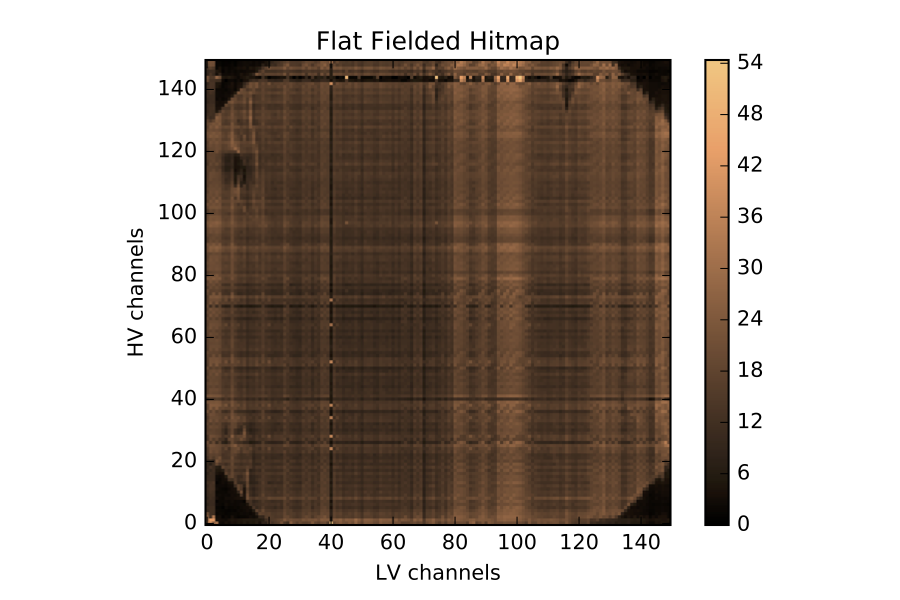} 
\caption{\label{fig:FF}Example of a 2D hitmap generated by matching coincident anode/cathode triggers from a Cs-137 source in the lab. The image was flat fielded with overall counts conserved to correct for various crystal effects. There are no counts in the corner regions (aside from spurious noise events) because the detector itself has rounded corners.}
\end{figure}

Each 3D-GeD is fabricated from a single germanium crystal. Orthogonal electrode strips are used to apply a 500--750 V bias across the crystal. A photoelectron creates a cloud of electrons and holes within the detector, which drift to opposite detector faces under the applied bias. Figure~\ref{fig:XYZ} depicts how the site of photon deposition is determined. The location of the energy deposition in two dimensions is determined by which anode strip and which cathode strip receive charge. The location in the third dimension is determined by the difference between the arrival times for electrons and holes on the opposite detector faces\cite{Amman2000, Amrose2003}. A 2D pixel map can be generated over many depositions to show the level of non-uniformity of charge collection (Figure~\ref{fig:FF}). 

%Figure~\ref{fig:XYZ} depicts how the site of photon deposition is determined. At left is a cartoon image showing how anode/cathode triggers are matched to determine a 2D position. At right are two plots showing a histogram of anode/cathode trigger time differences, with number of events plotted along the vertical axis. The 59.5keV photons for the Am-241 source is peaked to the left indicating absorption close to the surface, as expected for this relatively low energy source. The bottom plot shows timing differences for a Cs-137 source, which is dominated 662keV photons in the GRIPS energy range. These higher energy photons penetrate more uniformly into the crystal. The enhanced counts to the incident surface is from lower energy continuum photons which also come from the Cs-137 source. Assuming surface deposition for the 59.5keV photons, a .04mm/ns diffusion velocity @ 500V can be inferred. 

\begin{figure} [!h]
\centering
\includegraphics[scale=0.6]{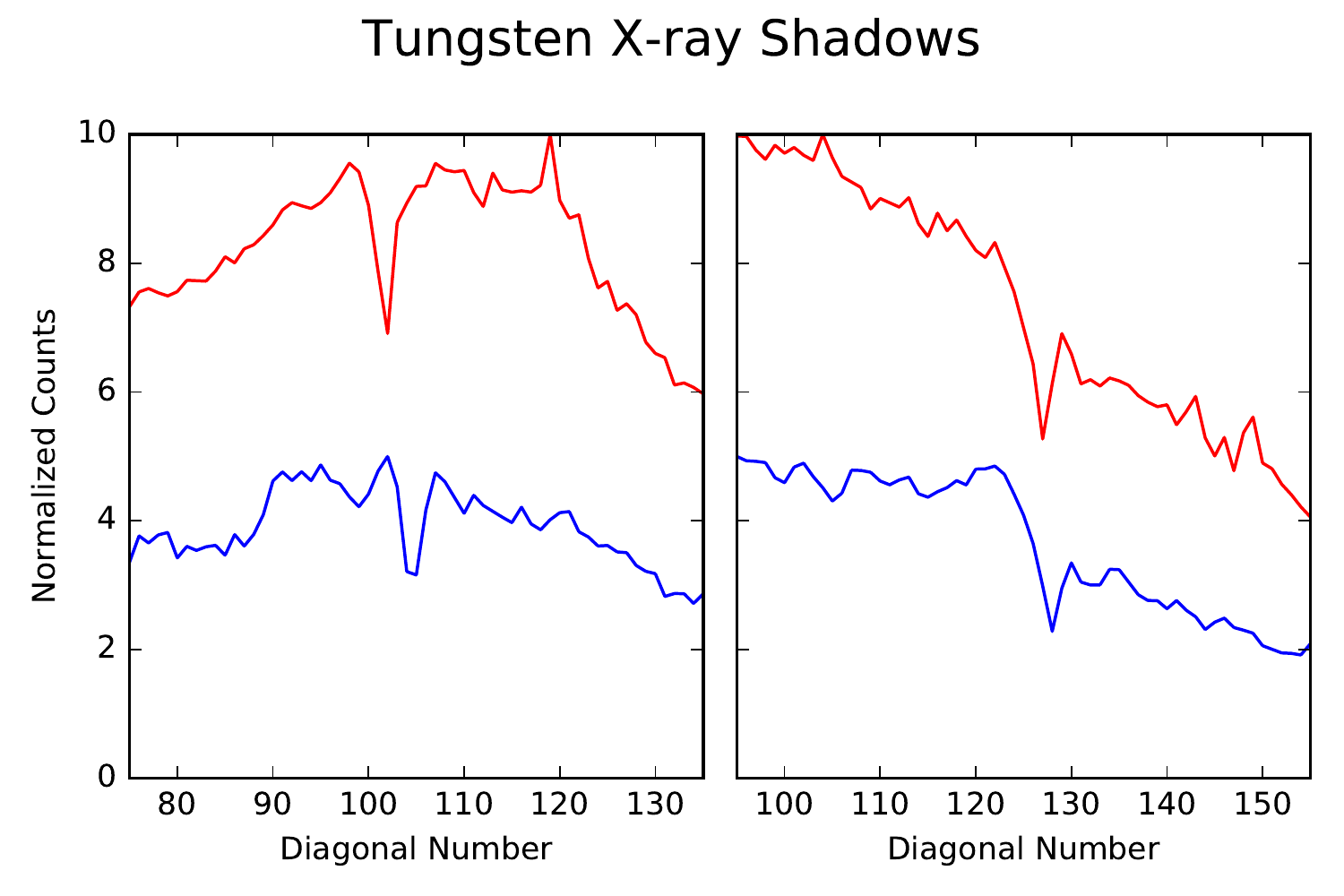} 
\caption{\label{fig:fid} Crossed tungsten rods were used to cast X-ray shadows along the detector diagonals. These plots show a the sum of counts along the diagonals for an Am-241 source in the lab, normalized to show two different detectors (D0 in red and D4 in blue) in single plot. The rods can be seen as dark "X"'s in Fig \ref{fig:aspect_target}. The quantity of pixels in a diagonal decreases with increasing diagonal number, giving the plots an overall decreasing count trend. The smaller count fluctuations are primarily due to variations in the energy thresholds for each pixel. }
\end{figure}

To verify the spatial resolution of the detectors and to precisely determine the positions of the detectors within the cryostat, crossed tungsten rods embedded into the cryostat top cast X-ray shadows onto the detectors. The spatial resolution was determined by measuring the width of this shadow in comparison to electrode pitch and rod diameter (0.46 mm). Figure~\ref{fig:fid} shows counts summed along the detector diagonals for two flight detectors. The dips at pixels 125 and 175 indicate the location and width of the tungsten rod shadow, showing that it is approximately the width of one strip pitch.  In cases of charge sharing, the extent of charge sharing between two neighbor strips could be used to determine sub-strip level position accuracy. GRIPS’s timing difference resolution of 50 ns FWHM translates into roughly $\sim$1 mm FWHM depth resolution within the crystal. These calibrations indicate a 0.5$\times$0.5$\times$1 mm voxel size for determining photon interactions sites within the GRIPS detectors.

%%%%%% -----------------------------------------------------------------------------------
\section{Spectra}
\label{sec:spectra}

%Nicole
%(1 page)

We provide here a preliminary evaluation of the performance of GRIPS's six flight detectors, with a detailed report to be provided in a forthcoming calibration paper.  GRIPS's analysis is particularly challenging due to its $\sim$1,800 detector channels that must be individually calibrated.  A complete description of the signal readout chain, including the ASIC electronics, can be found in an earlier manuscript\cite{Duncan2013}.

%Each of GRIPS's six instrumented flight detectors were equipped with 8 ASICs to provide a dedicated electronics chain for each of the 1,800 detector channels. In a forthcoming detector paper we will give a detailed report on channel calibrations and spectra in flight, providing here a general overview of performance and methods.

Preliminary line-width measurements show a typical energy resolution of 5 keV FWHM at 59.5 keV, but we are still refining our approach to subtracting common-mode noise.  Due to the fine strip pitch and dense circuit traces, there is significant correlated noise between multiple channels that can be quantified and removed.  This common-mode subtraction is implemented in post-processing of the raw data for each individual triggered event and has been extremely successful, reducing line-widths by 10s of keV.

The dominant source of common-mode noise is low-frequency noise resulting from the mechanical cryocooler and its associated power supply.  The ASIC has limited low-frequency filtering capabilities, and it was not feasible to introduce high-pass filtering in the analog-signal chain. When the cryocooler was briefly turned off, we were able to achieve line widths comparable to that previously reported for the GRIPS signal readout chain in a test cryostat cooled via liquid nitrogen: 3.5 keV FWHM at 59.5 keV\cite{Duncan2013}.

%We are expecting the linewidths to improve as we further develop our common-mode subtraction technique, which corrects for large scale baseline shifts that are correlated between multiple channels. These detectors achieved a 3.5 keV FWHM at 59.5 keV on the benchtop in a single detector cryostat which was cooled via liquid nitrogen\cite{Duncan2013}. 

\begin{figure} [!h]
\centering
\includegraphics[scale=0.65]{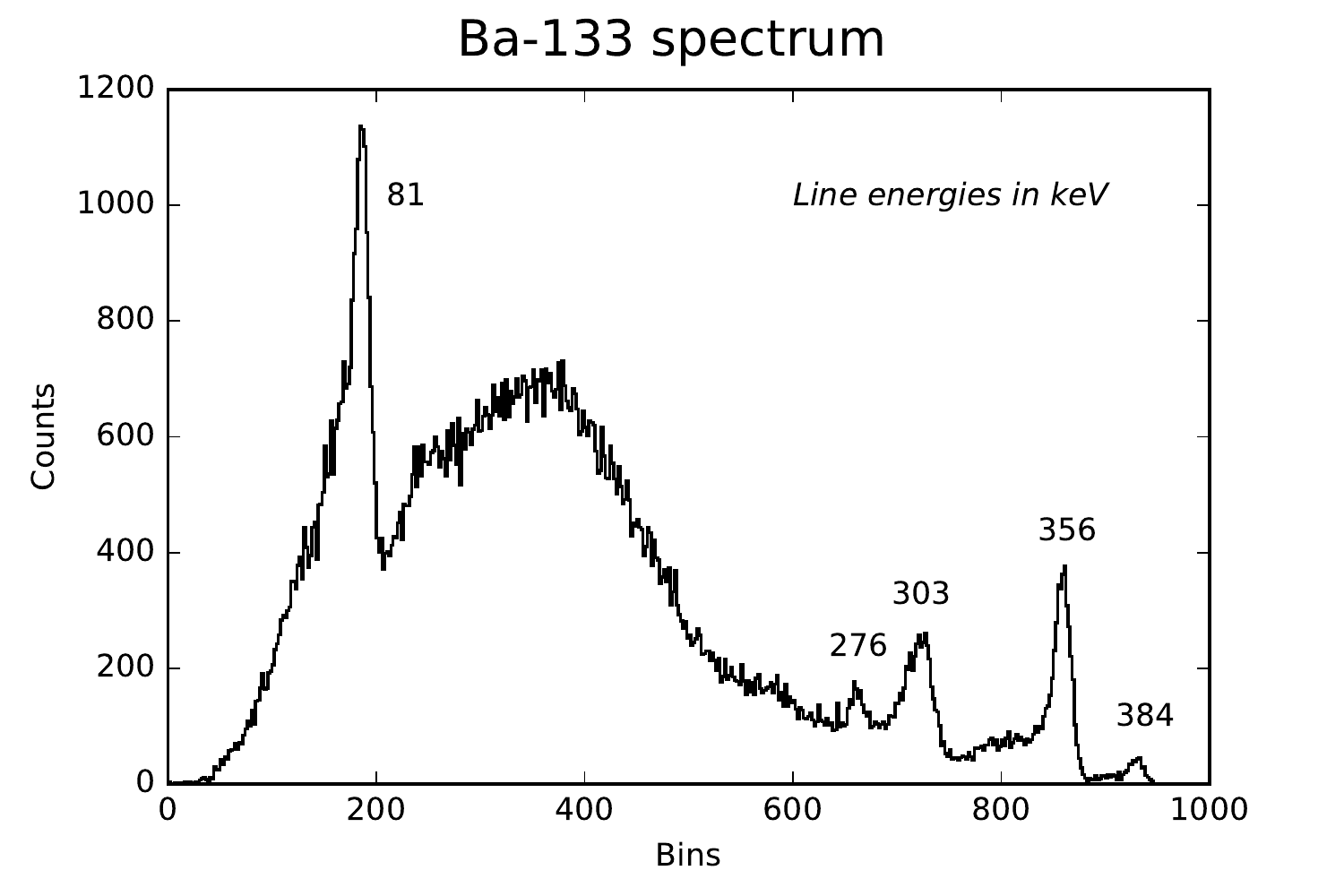} 
\caption{\label{fig:Ba}Single-strip spectrum for a barium-133 source, plotted in raw bin space. The low-energy threshold is set to $\sim$40 keV in this dataset. Incomplete charge collection of higher-energy events creates the wide bump-like feature around bin 375.}
\end{figure}

%The increased linewidth for the detectors in the spectrometer vs on the benchtop is primarily due to low frequency noise introduced by the mechanical cryocooler. Since the ASICs were manufactured before the flight cryostat was assembled and tested, there was limited ability to filter out this noise in the analog signal processing chain. This noise is predominately low frequency and our method of common mode subtraction has been extremely successful, reducing raw data linewidths by 10s of keV. Common mode subtraction is implemented in ground based, post-flight processing of the raw data for each individual triggered event.

\begin{figure} [!h]
\centering
\includegraphics[scale=0.6]{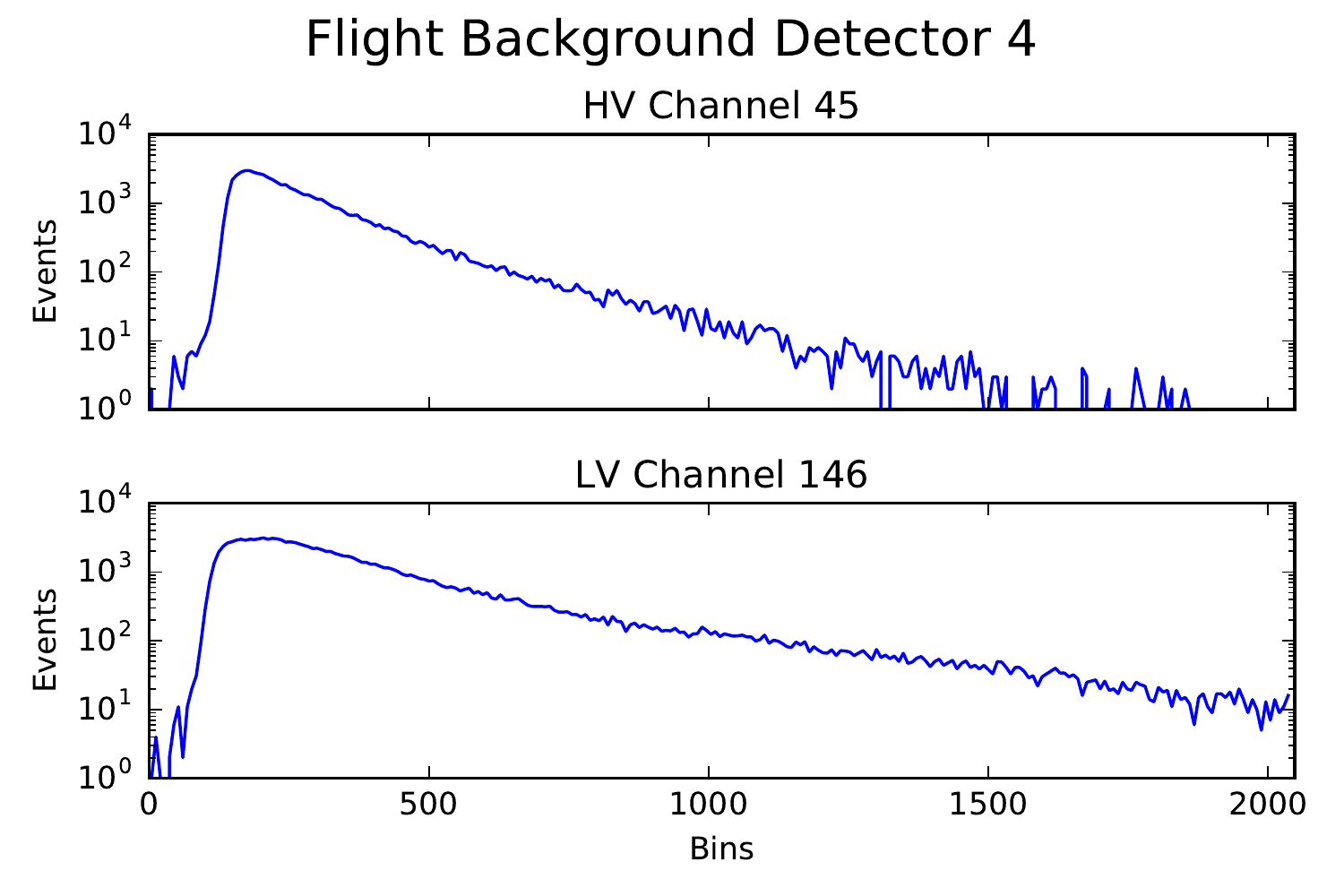} 
\caption{\label{fig:background}Antarctic background for an anode and cathode strip on Detector 4 integrated over 52 hours in flight. The spectra look as expected, peaking around the lower energy threshold and smoothly decreasing.}  
\end{figure}

The $\sim$30 keV low-energy spectral cutoff was determined by a combination of triggering thresholds, set to minimize spurious triggers, and the atmospheric energy cutoff, determined by the payload altitude and solar elevation in the sky. Due to the size of the ionization clouds and associated effects, there are not significant single-strip energy depositions above ~$\sim$1 MeV. The full 10 MeV energy range is achieved by summing over multiple strips in a scatter track. Four types of radioactive sources (Am-241, Co-60, Ba-133 and Cs-137) were used to determine the detector energy response. Figure \ref{fig:Ba} shows an example barium-133 single strip spectrum acquired as part of ground calibration. Figure~\ref{fig:background} shows a preliminary Antarctic background spectrum integrated over 52 hours in flight.

%%%%%% -----------------------------------------------------------------------------------
\section{Pointing Control System}
\label{sec:pointing}

%\subsection{Description}

GRIPS's pointing control system relied on a Sun sensor consisting of a quad-cell photodiode, with a 3.5 cm tube ending with a 5 mm square aperture mounted in front of it. The differences in voltage between photo-diode quadrants yields the accurate position of the Sun, down to a resolution of about 2 arcmin, corresponding to the smallest increment of the 11-bit ADC of micro-controller input. The Sun sensor had an 8$^{\circ}$ primary field of view (FOV), where the direction and distance of the Sun with respect to the Sun Sensor axis can be determined, and a secondary FOV extending to 24$^{\circ}$. When in the secondary FOV, it can be determined in which direction the Sun is, but not how far it is from the optical axis.

%The Absolute Encoder, attached to the elevation axle, provides a noisy reading of the boom elevation, while the Incremental Encoder, attached to the elevation motor (a linear actuator), provides much more accurate readings, but has no memory. Upon startup of the Pointing Control System, several Absolute Encoder readings are averaged together and used to initialize the Incremental Encoder, the default input to the elevation control loop.

%\subsection{Control principles}

Elevation was controlled by a linear actuator that would push/pull the boom as needed, and the position was measured by both an incremental encoder and an absolute encoder. The default elevation control logic is a simple "PD" (proportional \& derivative) loop. An alternate elevation control logic used a simpler constant-velocity ($\sim$0.1$^{\circ}/s$) control loop. The target elevation is computed by calculating the solar elevation at the time and location provided by GPS and correcting for differences with the Sun sensor output (accumulated over five minutes).

%To compensate for mechanical misalignments, an angular offset could also be added in the control loop. We have also implemented a more straightforward system, employing the Absolute Encoder when outside the FOV (moving very slowly to compensate for the noisy readings), and a simple constant-velocity ($\sim$0.1$^{\circ}/s$) control loop when the sun was within the primary FOV.

Azimuth was controlled by a motor attached to the main rotator, with no direct feedback of the position of the motor. The azimuth control logic is a classical PID control loop, and the "X" component of the Sun Sensor is the sole input. This control loop operates mostly independently from the elevation control loop, and begins seeking the Sun at a speed of about 0.5$^{\circ}$/s once the boom is at the proper elevation.  Once the Sun enters the FOV, the azimuth control loop adjusts the motor as needed to maintain tracking of the Sun.

The pointing control system functioned extremely well in flight. An angular offset (azimuth + elevation) below 0.5$^{\circ}$ rms was the requirement (driven mostly by imaging), and an average of 0.2$^{\circ}$ rms was achieved. 
While solar tracking performed well, acquiring the Sun in the first place proved to be difficult. We concluded that the Sun Sensor's FOV was too small for the whole system to properly damp a large-amplitude oscillation in azimuth ($\sim\pm$90$^{\circ}$). As a result, it took longer than expected to acquire the Sun during ascent, and it took $\sim$5 hours to re-acquire the Sun partway through the flight following a commanded reboot of the flight computer, fortunately with no loss to science.
Ground testing did not reveal this difficulty due to the limited azimuthal movements available in enclosed (wind-free) facilities.

\section{Aspect System}
\label{sec:aspect}

%Albert

%(2 pages)

The GRIPS aspect system provides the high-cadence, high-resolution pointing knowledge necessary for image reconstruction, thus allowing for the relaxed pointing-control requirements and mechanical requirements (e.g., twist).  The aspect system makes three primary measurements: the pitch/yaw offset of the optical bench to Sun center, the orientation of the MPRM, and the absolute roll of the optical bench.   Compared to earlier descriptions of the aspect system\cite{Duncan2013}, the final aspect system and its testing incorporated lessons learned from the Solar Aspect System (SAS) of the HEROES balloon instrument\cite{Christe2014}, whose design was adapted from the original GRIPS design.  Furthermore, the HEROES project graciously contributed components for the roll sensor, as discussed below.

\begin{figure} [!h]
\centering
\parbox{1.5in}{\includegraphics[width=1.5in]{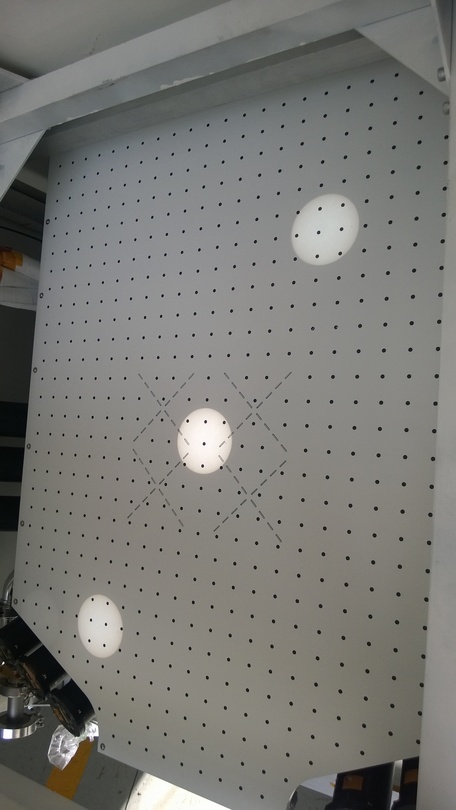}}%
\parbox{4in}{\includegraphics[width=4in]{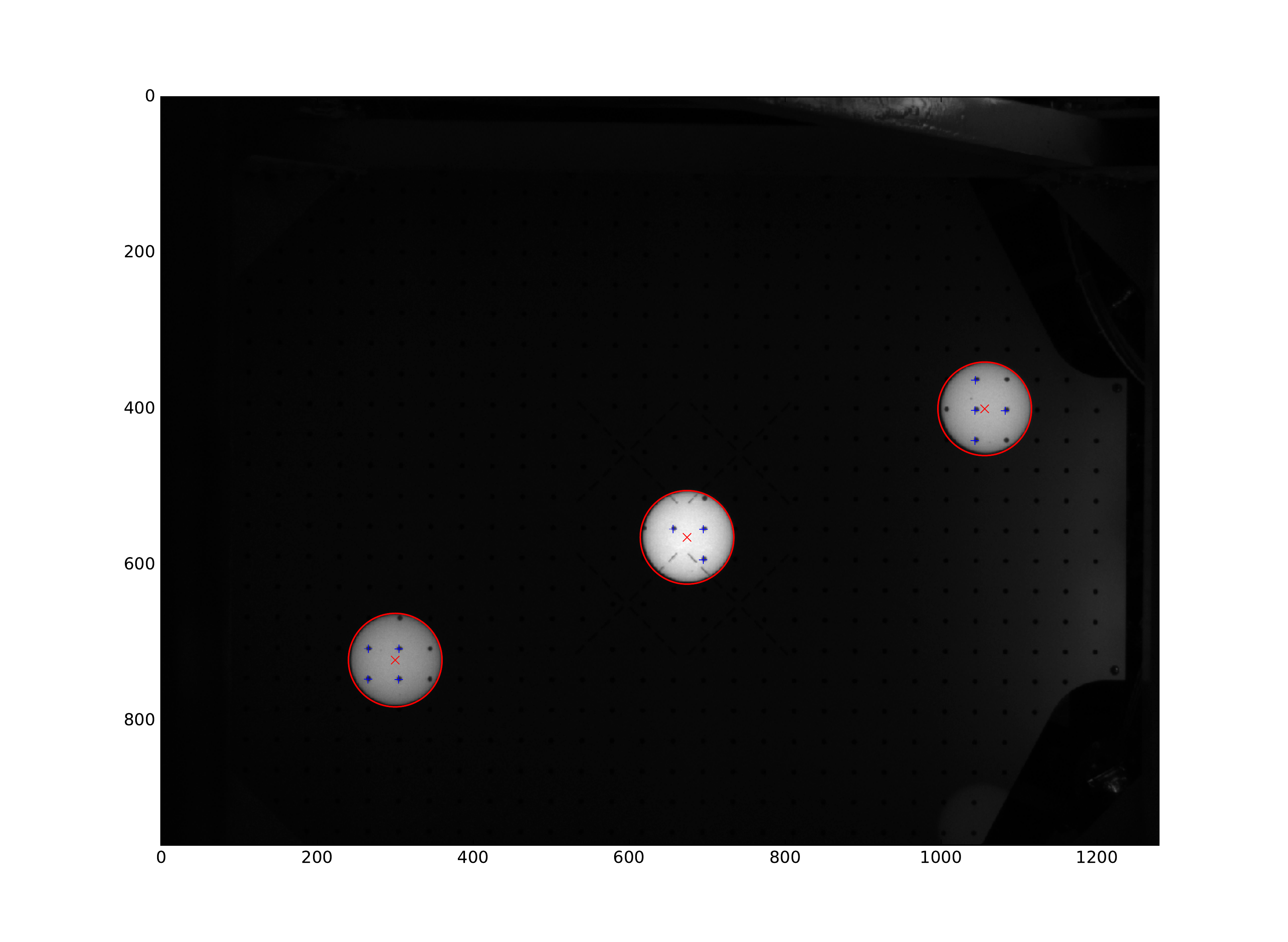}}%
\caption{\label{fig:aspect_target}\emph{Left}: Photo of the aspect target during a pre-flight pointing test showing the three images of the Sun, the fiducial pattern, and the "$\times$" patterns that are positioned above each stack of 3D-GeDs.  The Sun images in the lower left and in the upper right are at 90\% and 80\%, respectively, of the brightness of the central Sun image.  As the MPRM rotates, the outer Sun images appear to orbit around the central Sun image.\\
\emph{Right}: In-flight image from the pitch/yaw subsystem, with locations of the Sun (red $\times$ and circles) and illuminated fiducials (blue crosses) determined via a convolution filter.  The right edge of the frame corresponds to the nadir direction, and shows faint illumination from incompletely blocked Antarctic albedo.  Note also that the black tungsten rods, aligned with the "$\times$" patterns, are illuminated by the central Sun image.}
\end{figure}

The pitch/yaw offset and the MPRM orientation are measured simultaneously in a single subsystem.  This subsystem uses three 3-cm diameter plano-convex BK-7 lenses (focal length of 7780 mm) mounted within the MPRM assembly (Figure \ref{fig:mask}, right) to project three images of the Sun onto an aspect target mounted to the top of the detector cryostat (Figure~\ref{fig:aspect_target}, left).  Analysis of images from a 1280$\times$960 pixel CCD camera (Prosilica GT1290) pointed at the aspect target provides the locations of each of the images of the Sun relative to the fiducials in the aspect target, and thus relative to the center of the cryostat.  Because of this implementation, there are no stringent alignment requirements on this camera beyond including the aspect target within the camera FOV.

The pitch/yaw offset of the optical bench to Sun center is determined by the position of the central image of the Sun relative to precise fiducials (Figure \ref{fig:aspect_target}, right), which requires the pointing control system to point within a few degrees of the Sun.  With the effective plate scale of 16 arcsec per pixel, the pitch/yaw knowledge accuracy is better than a few arcsec.  The MPRM orientation is determined by the line of the three images of the Sun, and the two outrigger lenses have neutral density filters that reduce the intensity of the outrigger images to 90\% and 80\% of the central image.  Even during nominal pointing, as the MPRM rotates at $\sim$10 rpm, one of the outrigger images may fall beyond the aspect target and outside the camera FOV, but the MPRM orientation can still be determined by the remaining two images of the Sun.

To avoid the speckling observed in the HEROES SAS images that resulted from partial glossiness of screen-printing inks, the GRIPS implementation took advantage of the top surface of the cryostat immediately below the aspect target: the cryostat's top surface was painted flat black, the aspect target was painted flat white, and the desired high-contrast fiducials were achieved via precision-machined holes in the aspect target that revealed the black cryostat top.  The HEROES SAS experience also showed that the expanding grid of fiducials added an unnecessary level of complication as part of a mostly stable system with the entire aspect target within the camera FOV, so GRIPS used a simpler grid of constant spacing between its fiducials.

\begin{figure} [!h]
\centering
\includegraphics[width=4in]{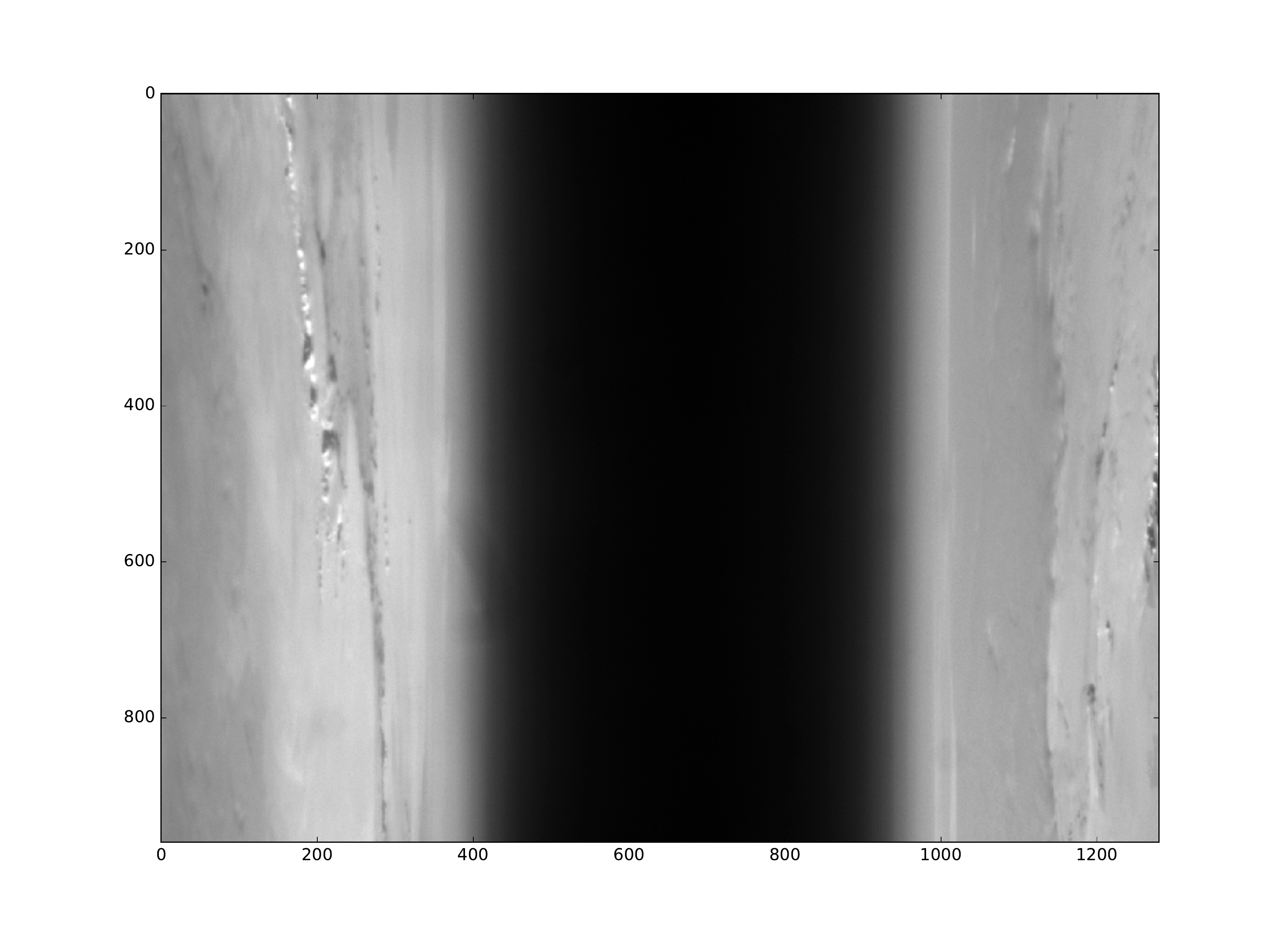} 
\caption{\label{fig:horizons}In-flight image from the roll sensor showing the opposing horizon views and the Antarctic surface. The axes indicate pixel number.}
\end{figure}

The absolute roll of the optical bench is determined by a separate subsystem from the above.  The roll sensor uses a silvered, knife-edge, right-angle prism to combine opposing views of the Earth horizon into a single image (Figure~\ref{fig:horizons}), as recorded by a 1280$\times$960 pixel CCD camera (also a ProSilica GT1290).  By measuring the locations of the opposing horizons, the sensor can determine absolute roll continuously indefinitely without any measurement drift as can be experienced by gyroscopes.  With the plate scale of 1.1 arcmin per pixel, the roll knowledge accuracy is better than 15 arcsec.  The HEROES project contributed the physical assembly and the mirrored prism, with the primary change being the exchange of the CCD camera.

To meet the science requirements for pointing knowledge, each of the two cameras transferred images at 4 Hz to a dedicated computer running Ubuntu Linux.  This aspect computer provides the command interface for controlling the cameras, produces real-time telemetry, and records the images for post-flight analysis on three $\sim$1-TB solid state drives.  To able to record all images for a potential month-long flight, images were saved using a combination of Rice compression and row decimation to reduce the storage needs with minimal impact to science.

Pre-flight calibration of the pitch/yaw subsystem consisted of a series of pointing tests when conditions in Antarctica were appropriate.  The wedge angles of each of the three lenses were calibrated by individually rotating each lens through four 90-degree increments while keeping the other two lenses fixed.  Pre-flight calibration of the roll subsystem consisted of using a theodolite as a visual feature within the camera's FOV and measuring the precise optical angular position of the theodolite using the theodolite itself.  The association of the pixel coordinates of the theodolite and the corresponding precise optical angles over a dozen positions of the theodolite allows determination of the plate scale and distortion of the roll-subsystem camera.
	
The aspect system performed exactly as expected during flight, and a detailed description is deferred to a future publication.  In brief, the pitch/yaw subsystem verified the performance of the pointing control system.  The roll subsystem measured the expected variety of roll motions, including the 24-second harmonic motion, or pendulation, of the entire balloon-payload system with a peak-to-peak amplitude of $\sim$1 arcmin.

%%%%%% -----------------------------------------------------------------------------------
\section{Data Storage and Telemetry}
\label{sec:telemetry}

The GRIPS flight computer received data and telemetry from the major subsystems over Ethernet connections: the pointing control system, the aspect system, the shield electronics, the spectrometer readout electronics, and the power system.  The flight computer also generated its own telemetry, which included information retrieved over serial connections with the cryocooler controller and the CSBF Support Instrumentation Package (SIP).  All data and telemetry (aside from images from the aspect system) were formatted as UDP packets with a GRIPS-specific packet header to identify the subsystem and telemetry type.  These packets were stored redundantly across three $\sim$1-TB solid-state drives in the flight computer.

Housekeeping telemetry and limited amounts of the science telemetry were downlinked throughout the flight. For the first $\sim$20 hours of flight, L-band line of sight (LOS) operations were possible, providing a 1 Mbps data rate which facilitated critical monitoring of the payload health during ascent and downlinking $\sim$25\% of the raw data from the 3D-GeDs.  Once LOS was no longer possible, we used a combination of the Tracking and Data Relay Satellite System (TDRSS) network and the Iridium satellite network, each of which had effective maximum data rates of $\sim$85 kbps.  (Iridium OpenPort also allowed for a direct SSH connection into the flight network.)  At this data rate, only $\sim$1\% of the raw data from the 3D-GeDs could be downlinked.  Instead, we relied on ``quicklook'' spectra and images generated by the flight computer to assess the detector health and performance during the flight.

\section{Future Directions}
\label{sec:conclusion}

The GRIPS balloon instrument flew successfully over Antarctica this past January, and the next steps for GRIPS will depend on the state of the hardware after it is recovered from Antarctica.  As an alternative to a future re-flight of GRIPS as a long-duration balloon (LDB) payload, GRIPS can be re-designed to be an ultra-long-duration balloon (ULDB) payload, making use of NASA's super-pressure balloon.  The super-pressure balloon allows for flights of several months in duration, albeit with the near-certain chance that the payload and its data vaults would not be recovered.  Detailed analysis of data from GRIPS's recent flight will guide the development of data-reduction techniques so that all of the science information can be retrieved through telemetry links alone.  Recurring flights of LDB or ULDB GRIPS payloads during the active years of the solar cycle would ensure that many large gamma-ray flares are observed.

A future space-based solar instrument can also make use of GRIPS's key technologies.  Given the ongoing progress in deployable booms, one can envision a spacecraft with scaled-up versions of the MPRM and the 3D-GeD spectrometer, separated by 20 meters or more.  Such an instrument would have unprecedented sensitivity to solar gamma rays at a few MeV and unprecedented angular resolution of better than 5 arcsec, which would resolve the smallest structures expected for accelerated ions in flares.  Not only would such a mission provide complete answers about the fundamental role of accelerated ions in flares, it would undoubtedly uncover new mysteries that we cannot even currently imagine. 

\section{Piggyback instruments}
\label{sec:piggyback}
In addition to the primary instrument, the GRIPS gondola also carried three piggyback instruments, and two of these instruments had solar-science objectives related to GRIPS's objectives.  The solar instruments, SOLAR-T and SMASH, specifically mounted to the telescope boom to ensure that they would be continuously pointed at the Sun during the long-duration flight.

  \subsection{SOLAR-T}
  
  SOLAR-T (PI: Pierre Kaufmann) was built to observe the full-disk Sun at 3 and 7 THz with high sensitivity and sub-second time resolution. Golay cell detectors are placed at the foci of two 7.6 cm Cassegrain telescopes. Low-pass filters consist of rough surface primary mirrors and membranes followed by metal mesh band-pass filters and choppers. Its redundant data acquisition system included Iridium telemetry for monitoring during the flight.
  
  The SOLAR-T performed successfully during the GRIPS flight. Solar disk brightness temperatures were determined: 5300K at 3 THz and 4700K at 7 THz. Burst detection sensitivity was 1\% of the full solar disk level at both frequencies. One impulsive event was detected from AR14289 on January 28, 2016, at 3 and 7 THz, peaking at 12:12:10 UT, time-coincident with bursts detected at 0.2 and 0.4 THz by the ground-based SST in Argentina, in H$\alpha$ by HASTA also in Argentina, and in EUV by SDO. One EUV bright blob forms at the top of an arch, “falls”, and hits the footpoint precisely at the same time of the THz impulsive burst. Fluxes at sub-THz and THz bands increase with frequency at both ranges, with a possible break in intensity somewhere between 0.4 and 3 THz.

  \subsection{SMASH: The SwRI Miniature Assembly for Solar Hard X-rays}
  
  SMASH (PI: Amir Caspi) was a technological demonstration of a new miniaturized hard X-ray (HXR) detector for future use on CubeSats and other small spacecraft, including the proposed CubeSat Imaging X-ray Solar Spectrometer (CubIXSS \cite{caspi16}).  SMASH demonstrated the space-borne application of the commercial off-the-shelf Amptek X123-CdTe, a miniature cadmium telluride photon-counting HXR spectrometer.
  %With modest resource requirements ($\sim$1/8 U, $\sim$200 g, $\sim$2.5 W) and low cost ($\sim$10K), the X123-CdTe is an attractive solution for HXR measurements from budget- and resource-limited platforms such as CubeSats. The detector has a physical area of 25 mm$^2$ and 1 mm fully-depleted thickness, with a $\sim$100 micron Be window; with on-board thermoelectric cooling and pulse pile-up rejection, it is nominally sensitive to solar photons from $\sim$5 to $\sim$100 keV with $\sim$0.5--1.0 keV FWHM resolution. For stratospheric flight, the effective low-energy limit was $\sim$20 keV due to atmospheric attenuation.
  The detector has a physical area of 25 mm$^2$ and 1 mm fully-depleted thickness and is nominally sensitive to solar photons from $\sim$5 to $\sim$100 keV with $\sim$0.5--1.0 keV FWHM resolution, but for the GRIPS flight, the effective low-energy limit was $\sim$20 keV due to atmospheric attenuation.

  Photons are accumulated into histogram spectra with customizable energy binning and integration time (cf. \cite{caspi15}); for SMASH, we used 1024 bins ($\sim$0.1 keV/bin) and 10 s integrations.  SMASH flew two identical detectors for redundancy and increased collecting area. The supporting electronics (power, CPU) were largely build-to-print using the Miniature X-ray Solar Spectrometer (MinXSS; \cite{mason16}) CubeSat design. Data was stored on board and also telemetered to the GRIPS flight computer for selective downlink to the ground.
  
  All of the observed C-class flares during the mission were below SMASH's sensitivity requirements for stratospheric flight, but a space-borne application would be sensitive to these smaller flares. Analysis of the microburst periods is pending. Crucially, however, the X123 detectors operated nominally during the entire $\sim$12-day flight, with no anomalies or upsets recorded, thus proving the operation of these detectors in a near-space environment.

  \subsection{TILDAE: Turbulence and Intermittency Long-Duration Atmospheric
Experiment}

The Turbulence and Intermittency Long-Duration Atmospheric Experiment [TILDAE: Co-PI’s: Raffaele Marino (NCAR, ENS Lyon) and Bennett A. Maruca (U.
Delaware, UC Berkeley)], encompassed a sonic anemometer (manufactured by Applied Technologies, Longmont, CO) to measure the propagation time of sound waves, to infer wind velocity and air temperature at a 200 Hz cadence \cite{kaimal1978}. TILDAE was designed to observe the intermittent structures (including "bursts" of turbulence) that develop in the stratification layers of the lower and mid stratosphere \cite{theuerkauf2011}. Its flight on GRIPS was one of the first attempts to use this type of anemometer in the stratosphere, taking advantage of the gondola's long-duration flight, gradual changes in altitude, and orientation (azimuth) control. In general, sonic anemometers have good sensitivity and excellent calibration \cite{cuerva2000}.

%%%%%% -----------------------------------------------------------------------------------
\acknowledgments    

The GRIPS project was funded in large part by NASA grants NNX08BA28G and NNX13AJ21G.  N. Duncan's work was partially supported by a NASA Graduate Student Researchers Program (GSRP) fellowship.  B. Maruca's work on both GRIPS and TILDAE was supported by the Charles Hard Townes Postdoctoral Fellowship.  TILDAE was also supported by the University of Calabria via the Turboplasmas project (Marie Curie FP7 PIRSES-2010-269297).  SMASH was supported by an internal R\&D grant at Southwest Research Institute.  SOLAR-T was supported by multiple grants from S\~{a}o Paulo Research Foundation (FAPESP).

%%%%%% -----------------------------------------------------------------------------------
% References
%\cite{Alred03}
\bibliography{report} % bibliography data in report.bib
\bibliographystyle{spiebib} % makes bibtex use spiebib.bst

\end{document}